\documentclass[twocolumn]{aastex63}
\usepackage{graphics,epsf}
\usepackage{amsmath}                % American Mathematical Society package
\usepackage{amsfonts}               % American Mathematical Society fonts
\usepackage{amssymb}                % American Mathematical Society symbol
\usepackage{epsfig}                 % EPS figures
\usepackage{appendix}
\usepackage{graphicx}
\usepackage{float}
\usepackage{color}
\usepackage{multirow}
\usepackage{colortbl}
\usepackage[para,online,flushleft]{threeparttable}

\newcommand{\m}{{~\rm m}}

\newcommand{\s}{{~\rm s}}

\newcommand{\g}{{~\rm g}}
\newcommand{\G}{{~\rm G}}
\newcommand{\K}{{~\rm K}}
\newcommand{\erg}{{~\rm erg}}
\newcommand{\yr}{{~\rm yr}}

\begin{document}

\title{Onset of common envelope evolution during a core helium flash by rapid envelope expansion}

%\correspondingauthor{Ealeal Bear, Noam Soker}
%\email{}

\author{Stanislav Fainer}
\affiliation{Department of Physics, Technion – Israel Institute of Technology, Haifa 3200003, Israel; ealealbh@gmail.com; soker@physics.technion.ac.il}
  
\author{Ealeal Bear}
\affiliation{Department of Physics, Technion – Israel Institute of Technology, Haifa 3200003, Israel; ealealbh@gmail.com; soker@physics.technion.ac.il}

\author[0000-0003-0375-8987]{Noam Soker}
\affiliation{Department of Physics, Technion – Israel Institute of Technology, Haifa 3200003, Israel; ealealbh@gmail.com; soker@physics.technion.ac.il}

\begin{abstract}
We suggest that the vigorous core convection during core helium flash on the tip of the red giant branch (RGB) of low mass stars excites waves that carry energy to the envelope and inflate it for few years to increase the number of extreme horizontal branch (EHB; sdB and sdO) stars with masses of $\simeq 0.47 M_\odot$   with respect to canonical binary evolution.  
Using the open-source \textsc{mesa-binary} we follow the evolution of a number of eccentric binary systems with an initial primary stellar mass of $1.6 M_\odot$. The energy that the waves carry to the envelope leads to envelope expansion at the tip of the RGB. The inflated RGB star engulfs many secondary stars to start a CEE that otherwise would not occur. If the secondary star manages to remove most of the RGB envelope the primary evolves to become an EHB star with a mass of $\simeq 0.47 M_\odot$.   However, we expect that in most  cases the secondary star does not have time to spiral-in to close orbits. It rather ends at a large orbit and leaves a massive enough envelope for the primary star to later evolve along the asymptotic giant branch and to engulf the secondary star, therefore forming a non-spherical planetary nebula.  
\end{abstract}

% ==========================================================
\section{Introduction} 
\label{sec:intro}
% ==========================================================

Hot sdB and sdO stars are extreme horizontal branch (EHB) stars with a very low  envelope mass (e.g., \citealt{Heber1986, Heber2016, Geieretal2019}). Their very low envelope mass, their total mass of $\simeq 0.47 M_\odot$ (e.g., \citealt{Luoetal2019, Corcoranetal2021, Schaffenrothetal2022}), and the helium burning in their core imply that their red giant branch (RGB) star progenitors lost most of their envelope on the tip of the RGB. Theoretical models for the formation of EHB stars consider mainly stellar binary interaction (e.g., \citealt{Hanetal2002, Hanetal2003, Geetal2022, Geieretal2022}). However, some scenarios involve also substellar companions (e.g., \citealt{Soker1998, NelemansTauris1998, Krameretal2020, Irrgangetal2021, Schaffenrothetal2021,VanGrooteletal2021}). Observations show indeed that some sdB stars have a close brown dwarf companion (e.g., \citealt{Schaffenrothetal2015, Daietal2022}). Despite decades of studies there are still some puzzles regarding their formation (e.g., \citealt{Pelisolietal2020}). 

In the present paper we concentrate on the common envelope evolution (CEE) channel, which is one of three stellar binary channels to form EHB stars (e.g., \citealt{Hanetal2002, Hanetal2003, Hanetal2020}). Observations of close companions to EHB (hot subdwarf) stars, mainly white dwarfs (WDs) and spectral type M main sequence stars (e.g., \citealt{Copperwheatetal2011, Geieretal2011, Kupferetal2020, Baranetal2021, Corcoranetal2021, Kruckowetal2021, Daietal2022}) support the CEE channel. 

Stars in the zero age main sequence mass range of $0.8 M_\odot \la M_{\rm ZAMS} \la 2 M_\odot$ terminate their RGB evolution with a core helium flash. In an earlier paper we \citep{Bearetal2021} proposed that the vigorous core convection that develops during the core helium flash excites waves that propagate to the envelope and deposit their energy in the outer envelope. The deposition of energy causes rapid envelope expansion \citep{Bearetal2021}. \cite{Merlovetal2021} study how this rapid expansion can bring the RGB star to engulf a planet companion during the core helium flash, just as the star is about to terminate its RGB evolution. In the present study we use the binary-\textsc{mesa} evolutionary code (section \ref{sec:Numerics}) to follow the evolution of a low mass companion to the RGB star under the same assumptions as \cite{Bearetal2021} and \cite{Merlovetal2021} made. 

In \cite{Bearetal2021} and \cite{Merlovetal2021} we based our scenario on the studies by \cite{QuataertShiode2012} and \cite{ShiodeQuataert2014} on the excitation of waves by the vigorous convection in the cores of pre-supernova massive stars.  
The vigorous convection can excite waves with a power that is about one per cent of the convective power (\citealt{QuataertShiode2012, LecoanetQuataert2013, ShiodeQuataert2014}). This small fraction of the power in the waves relative to the kinetic power of the convection, nonetheless, might cause large envelope expansion if the waves deposit their energy in the envelope  (e.g., \citealt{McleySoker2014, Fuller2017}). 
We also based our proposal on the finding of \cite{Mocaketal2010} that the core convection, which the core helium flash drives, excites gravity waves. \cite{Mocaketal2010} find also that the core convection carries most of the nuclear reactions energy.
We note that other possible roles of the waves that the core convection excites during the core helium flash might be extra mixing near the core-envelope boundary \citep{Schwab2020}, periodic photometric variability in hot subdwarf stars \citep{MillerBertolamietal2020},  and influencing the hydrogen burning \citep{JermynFuller2022}. 

 Let us detail the role of the waves. Although the core helium flash is a mildly explosive event most of the energy that the rapid nuclear helium burning releases goes to heat the core and then to cause its rapid expansion. Namely, the expanding core drains most of the energy that the explosive helium burning releases. For that, spherically symmetric stellar evolution simulations cannot find much envelope expansion, if at all. Only the propagation of non-radial modes, that are unattainable to spherical codes, might affect the envelope. The outbursts of some massive stars that occur years to weeks before explosion hint to the possibility of the propagation of convection-excited waves from the core to the envelope (e.g., \citealt{QuataertShiode2012}). Because core helium flashes also lead to vigorous convection that excite waves \citep{Mocaketal2010} we study the possible outcome of the energy that the waves might carry from the core to the envelope.

With a stellar evolutionary code (section \ref{sec:Numerics}) we study the expansion of the envelope that might lead to the engulfment of a low mass stellar companion, i.e., a low mass main sequence star or a WD (section \ref{sec:Evolution}). In section \ref{sec:Cumulative} we present the cumulative mass distribution that our results imply for EHB stars  together with lower mass helium WDs that are descendants of early termination of the RGB following a CEE.  In section \ref{sec:PNe} we discuss implications to the shaping of planetary nebulae that are descendants of low mass star. We summarize our study in section \ref{sec:Summary}.

% ==========================================================
\section{Numerical Scheme and assumptions}
\label{sec:Numerics}
% ==========================================================
% =====================
\subsection{\textsc{mesa}}
\label{subsec:Code}
% =====================
We follow the evolution of a stellar model with $M_{\rm ZAMS,1}=1.6M_\odot$ and metalicity of $z=0.02$, the primary star in a binary system, using the evolutionary stellar code Modules for Experiments in Stellar Astrophysics (\textsc{mesa}; \citealt{Paxtonetal2011, Paxtonetal2013, Paxtonetal2015, Paxtonetal2018, Paxtonetal2019}), version 10398 in its binary mode-point mass. Namely, we do not follow the evolution of the secondary star. We simulate cases with secondary stellar masses of $M_2=M_{\rm ZAMS,2}=0.3, 0.5M_\odot$.

We first follow an example from that version of \textsc{mesa} (\texttt{star~plus~point~mass}). In our simulations accretion is negligible and the secondary mass is practically constant along the evolution. Hence, we take \texttt{limit~retention~by~mdot~edd = .false.}. The primary star loses mass (with the parameter $\eta=0.7$ in numerical subroutine \texttt{1~pre~MS~to~WD}). We enable circularisation and synchronisation of the primary due to the tidal force (numerically we set \texttt{do~tidal~circ=true} and \texttt{do~tidal~sync=true}).
All other parameters are the defaults of \textsc{mesa}. 

Our condition to the onset of a CEE is that the periastron distance is smaller than the primary radius $a(1-e) < R_1$, where $a$ is the semi-major axis, $e$ is the eccentricity, and $R_1$ is the primary radius. 

% =====================
\subsection{\textsc{Numerical procedure}}
\label{subsec:Procedure}
% =====================

We start each simulation by assigning the secondary mass, which is constant along the evolution, $M_2=M_{\rm ZAMS,2}=0.3 M_\odot$ or $0.5 M_\odot$, the initial eccentricity $e_0$, and the initial semi-major axis $a_0$. In all cases that we present in this study the initial primary mass is $M_{\rm ZAMS,1}=1.6 M_\odot$. 

We perform some simulations without the deposition of the wave energy during the core helium flash ($E_{\rm wave}=0$), and some simulations with the deposition of wave energy to the envelope (see section \ref{sec:intro}). We base the energy deposition on the simulations of  \cite{Merlovetal2021}.
We take the power that the waves carry to the envelope to be $L_{\rm wave}=2 \times 10^4 L_\odot$ and deposit the energy for four years. The power is an intermediate value in the range of powers that \cite{Merlovetal2021} used, and the duration is the same as in all simulations of \cite{Merlovetal2021}. \cite{Merlovetal2021} deposited the energy into either the outer $20 \%$, $50 \%$, or $80 \%$ of the envelope mass. We here deposit the wave energy into the outer $50 \%$ of the envelope mass (for more details see that paper). 
We follow \cite{Bearetal2021} and start the four years of energy deposition such that it ends at the time of peak convection luminosity in the core. 

% ==========================================================
\section{Evolution towards a CEE}
\label{sec:Evolution}
% ==========================================================

We first evolve the RGB star without deposition of wave energy, i.e., $L_{\rm wave}=0$, and find the maximum value of the initial semi-major axis $a_{\rm 0,CEE}$ for which a CEE still takes place. The condition for the onset of a CEE is  $a(1-e) < R_1$ (section \ref{subsec:Code}). We present the results in Table \ref{tab:Table1}. In the first two columns we list the input parameters of the secondary mass $M_2$ and of the initial eccentricity $e_0$, respectively.
The typical uncertainty in the values of the maximum orbital separations is $\approx \pm 2 \%$. The uncertainty is because tidal forces determine this maximal orbital separation and they are very sensitive to the stellar parameters, in particular its radius. We present the values of $a_{\rm 0,CEE}$ in the third column of Table \ref{tab:Table1}. 
In the fourth column we list the maximum semi-major axis for which the binary system enters a CEE when we deposit the wave-energy during the core helium flash according to the procedure of section \ref{subsec:Procedure}, i.e., a power of $L_{\rm wave}=2 \times 10^4 L_\odot$ for four years which gives a total deposited energy to the envelope of $E_{\rm wave} = 9.7 \times 10^{45} \erg$. 
% TTTTTTTTTTTTTTTTTTTTTTTTTTTTTTTTTTTTTTTTTTTTTTTTT
\begin{table}[]
\centering
\begin{tabular}{|cc|c|c|}
\hline
\multicolumn{2}{|c|}{Initial parameters} & $L_{\rm wave}=0$ & $L_{\rm wave}=2 \times 10^4L_\odot$ \\ \hline
\multicolumn{1}{|c|}{{$M_2$}} & {$e_0$} & {$a_{\rm 0,CEE}$} & {$a_{\rm 0,CEE,W}$} \\ \hline
\multicolumn{1}{|c|}{{$M_\odot$}} & {} & {$R_\odot$} & {$R_\odot$} \\ \hline
\multicolumn{1}{|c|}{0.3} & 0 & 280 & 540 \\ \hline
\multicolumn{1}{|c|}{0.3} & 0.4 & 340 & 650 \\ \hline
\multicolumn{1}{|c|}{0.3} & 0.8 & 780 & 1550 \\ \hline
\multicolumn{1}{|c|}{0.5} & 0 & 310 & 580 \\ \hline
\multicolumn{1}{|c|}{0.5} & 0.4 & 370 & 700 \\ \hline
\multicolumn{1}{|c|}{0.5} & 0.8 & 860 & 1600 \\ \hline
\end{tabular}
\caption{The maximum semi-major axis for which the binary enters a CEE without energy deposition to the envelope $a_{\rm 0,CEE}$ (third column) and when we deposit wave-energy to the envelope (see section \ref{subsec:Procedure}) $a_{\rm 0,CEE,W}$ (fourth column). The typical uncertainties in these values are $\approx \pm 2 \%$. In the first column we list the secondary stellar mass $M_2$ and in the second column the initial eccentricity $e_0$. In all cases the initial primary mass is $M_{\rm ZAMS,1}=1.6 M_\odot$.  }
\label{tab:Table1}
\end{table}
% TTTTTTTTTTTTTTTTTTTTTTTTTTTTTTTTTTTTT

\begin{figure}%[ht]
 %\centering
 \vskip -5.0 cm
 \hskip -2.00 cm
 \includegraphics[scale=0.6]{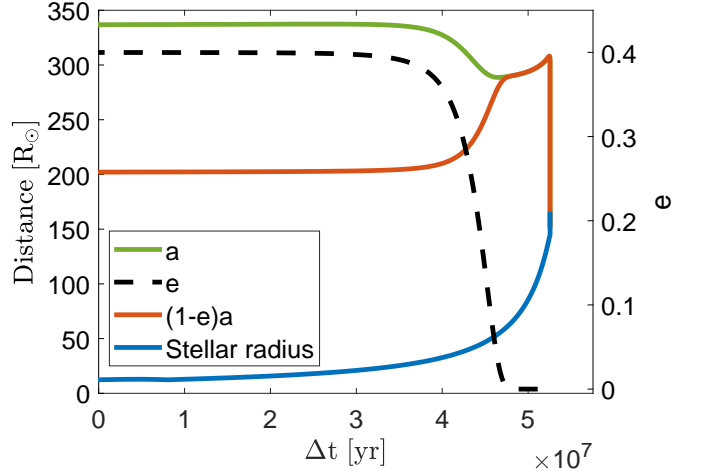}\\
 \vskip -5.00 cm
\caption{The semi-major axis $a$ (green line), the eccentricity $e$ (black-dashed line with a scale on the right vertical axis), the periastron distance $(1-e)a$ (red line) and the stellar radius $R_1$ (blue line), as a function of $\Delta t \equiv t-2.2\times 10^9\yr$ without energy deposition. Initial eccentricity, semi-major axis, and masses are $e_0=0.4$, $a_0=335R_\odot$, $M_{\rm ZAMS,1}=1.6M_\odot$, and $M_2=0.3M_\odot$, respectively. Note that the system enters a CEE at the last time of this graph. }
 \label{fig:CEE_NO_INJECTIONgraph}
 \end{figure}

\begin{figure}%[ht]
 %\centering
 \vskip -5.0 cm
 \hskip -2.00 cm
 \includegraphics[scale=0.6]{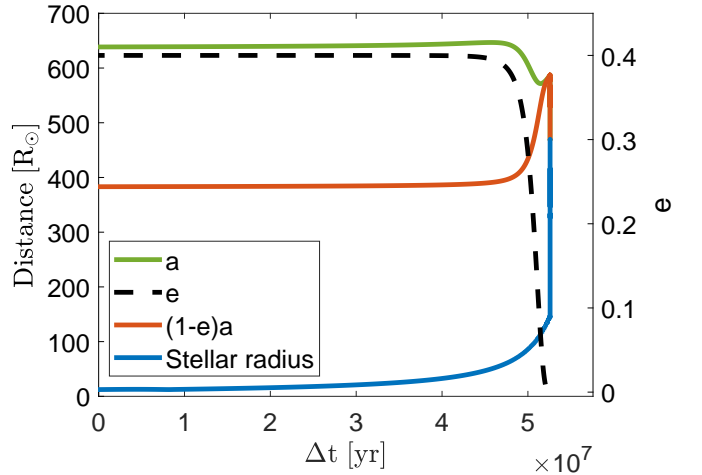}\\
 \vskip -5.00 cm
\caption{Similar to Fig. \ref {fig:CEE_NO_INJECTIONgraph} but now the initial semi-major axis is $a_0=635R_\odot$ and in the last years of evolution we deposit to the envelope the wave energy according to section \ref{subsec:Procedure}. }
 \label{fig:CEE_INJECTIONgraph}
 \end{figure}

In section \ref{sec:Cumulative} we will study in more details binary systems with the initial properties of $M_{\rm ZAMS,1}=1.6M_\odot$, $M_2=0.3M_\odot$ and $e_0=0.4$. In Fig. \ref{fig:CEE_NO_INJECTIONgraph} we present the evolution of the orbital parameters for a binary with these parameters and with an initial semi-major axis of $a_0=335R_\odot$, which is close to the maximum value for which CEE takes place without including wave energy for the above binary parameters (Table \ref{tab:Table1}). 
In Fig. \ref{fig:CEE_INJECTIONgraph} we show the evolution for the same binary masses and eccentricity, but now with the deposition of the wave energy to the envelope and for $a_0=635R_\odot$, which is close to the maximum value for which the system experiences a CEE for this value of wave energy deposition (Table \ref{tab:Table1}). 
% FFFFFFFFFFFFFFFFFFFFFFFFFFFFFFFFFFFFFFFFF
  %[ht]
% FFFFFFFFFFFFFFFFFFFFFFFFFFFFFFFFFFFFFF
% FFFFFFFFFFFFFFFFFFFFFFFFFFFFFFFFFFFFFFFFF
  %[ht]
% FFFFFFFFFFFFFFFFFFFFFFFFFFFFFFFFFFFFFF

From the results that we present in Table \ref{tab:Table1} we learn that there is a large range of initial semi-major axes for which binary systems enter a CEE, if waves propagate to the envelope and deposit energy there according to our assumptions (section \ref{subsec:Procedure}), but if such energy deposition does not take place then the binary systems avoid CEE. 

Let us elaborate on the onset of the CEE during the brief envelope expansion. We first point out that we assume (section \ref{subsec:Procedure}) that the energy deposition by waves causes envelope inflation. Namely, we neglect any mass ejection that this energy might cause. In doing so we follow the claim of \cite{McleySoker2014} for the case of pre-supernova wave energy deposition in massive stars. \cite{McleySoker2014} find that the extra wave pressure and wave-energy dissipation lead to envelope expansion rather than to major mass ejection. 

Our second point refers to two major differences from the onset of a regular CEE (during regular binary evolution). (1) The rapid expansion and contraction,  for several years, does not allow for a slow spiralling-in. Numerical simulations show that after engulfment the first spiralling-in phase (the plunge-in phase) proceeds on a timescale which is about the Keplerian time scale (e.g.,  \citealt{Reichardtetal2020, Sandetal2020, GlanzPerets2021, Lopezcamaraetal2022}). For our most distance orbital separations, $a \ga 500 R_\odot$ at engulfment, the orbital period is about equal or longer than the expansion time. In those cases the companion might spiral-in for a short distance and remove part of the envelope.  What might help is that the spiralling-in of the companion occurs because orbital energy is channelled to the RGB envelope. This extra energy further inflates the envelope, hence prolonging the engulfment time. We do not simulate this process as we terminate our simulations at the very beginning of the CEE. 
(2) On the other hand, the rapid expansion implies that the secondary star has no time to spin-up much the RGB envelope. In regular evolution the secondary star spins-up the envelope. The slower (than the orbital velocity itself) relative velocity of the secondary star inside the envelope reduces the tidal force before entering a CEE and the dynamical friction at the early CEE. In the case of rapid expansion that we study here, spin-up is negligible and therefore the tidal force and the dynamical friction inside the envelope are stronger than in regular binary evolution. Namely, the spiralling-in is even faster. 

We discuss possible implications for cases where the secondary star manages to remove most of the envelope in section \ref{sec:Cumulative}, and for cases where the secondary spirals-in to a limited distance in section \ref{sec:PNe}.

% ==========================================================
\section{Cumulative mass distribution}
\label{sec:Cumulative}
% ==========================================================

In this section we assume that after the secondary star enters a CEE it spirals-in inside the RGB envelope and removes all the envelope, leaving behind a binary system of a secondary main sequence star and a primary that is the remnant of the RGB star.  If the CEE takes place at an early RGB phase the remnant is a helium WD (He WD), while if the CEE takes place on the tip of the RGB the remnant is an EHB star that later evolves to become a CO WD.  In reality we expect that if the secondary enters a CEE in a regular RGB evolution then in all cases it indeed removes most of, or even the entire, envelope (if it is massive enough). However, in cases where the secondary star enters the envelope during the brief (few years) envelope expansion in some cases it would not have sufficient time to spiral-in before the envelope contracts back. Therefore, the secondary star removes only a fraction of the envelope (see discussion in section \ref{sec:Evolution}). 

In this section we consider the implications to the cumulative mass distribution of  WD remnants with masses of $M_{\rm WD} \la 0.47 M_\odot$. These WDs that result from RGB CEE (RGCE) that removes their entire hydrogen-rich envelope on the RGB are of two kinds. Either CO WD remnants of EHB stars, or He WD remnants of CEE on the RGB before helium ignition. The EHB will turn to CO WDs of mass $\simeq 0.47 M_\odot$, while the He WDs will cool as He WD of lower masses.   In section \ref{sec:PNe} we will discuss implications to planetary nebulae (PNe). 

In estimating the number of WDs that are descendant of CEE on the RGB per unit mass range we assume first that once the RGB star engulfs a stellar companion within a very short time compared to the time the envelope might shrink by itself, the companion spirals-in and removes the entire RGB envelope. 
We need the relative number of stellar companions as function of initial semi-major axis $a_0$ and the mass of the RGB core when the companion and the RGB star start their CEE. 

We take the relative number of solar-like star binaries per orbital period from \cite{MoeDiStefano2017}, i.e., the fitting in their figure 31 (also red line in the lower panel of their figure 37). We make an approximation to this line as
\begin{equation}
 f \equiv \frac{ d N}{d \log P({\rm day}) } \simeq 0.02 \log P(\rm day). 
\label{eq:MoeDiStefano}    
\end{equation}
For our binary systems with $M_1=1.6 M_\odot$ and $M_2=0.3 M_\odot$ this relation for the semi-major axis reads
\begin{equation}
\frac{ d N}{d \log a_0(R_\odot) } 
% \simeq 0.02 \frac{3}{2}   \left[ 2.42 + 1.5 \log a({\rm AU})\right] . 
%  \simeq 0.0727 +0.045 \log a({\rm AU}) 
   \simeq 
   0.045 \log a_0(R_\odot)  - 0.03225 .
\label{eq:dNda}    
\end{equation}
We will use the number of systems $N$ as a relative number, so its scaling is of no concern here. 

\begin{figure}%[ht]
 %\centering
 \vskip -5.0 cm
 \hskip -2.00 cm
 \includegraphics[scale=0.6]{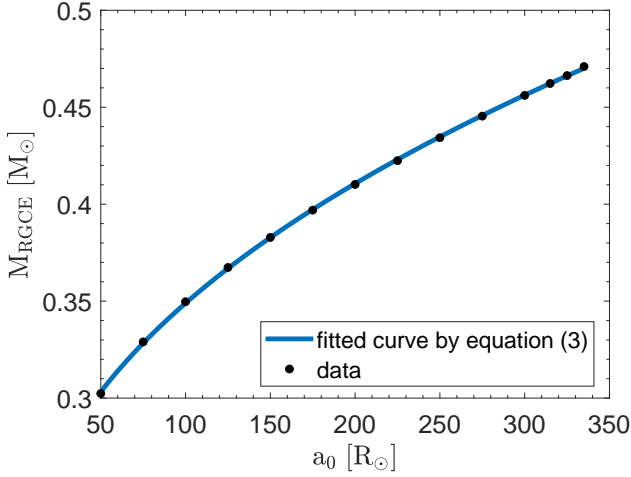}\\
 \vskip -5.00 cm
\caption{The black dots represent the RGB core mass when CEE starts on the RGB, for different initial orbital separations $a_0$.  We assume that the core mass is the final mass of the WD that is the descendant of the CEE on the RGB, $M_{\rm RGCE}=M_{\rm core}$.   In all cases $M_{\rm ZAMS,1}=1.6M_\odot$, $M_2=0.3M_\odot$ and $e_0=0.4$, and there is no wave energy deposition. The blue curve is our fit to the data according to equation (\ref{eq:dMdA}).}
 \label{fig:CEE_CORE_MASS_noEFFECTgraph}
 \end{figure}

We will demonstrate the derivation of the expression for one set of initial masses and eccentricity. Other cases are qualitatively similar. 
In Fig. \ref{fig:CEE_CORE_MASS_noEFFECTgraph} we present the mass of the core of the RGB star with a zero age main sequence mass of $M_{\rm ZAMS,1}=1.6 M_\odot$ when it engulfs a companion of mass $M_2=0.3 M_\odot$ as function of the initial semi-major axis $a_0$ of the system. In all these systems the initial eccentricity is $e_0=0.4$, and there is  no wave energy deposition. As stated, we assume that the final mass of the WD, $M_{\rm RGCE}$, equals the core mass at the onset of the RGB CEE, i.e., $M_{\rm RGCE}=M_{\rm core}$  (whether a He WD or a CO WD descendant of an EHB star). 
% FFFFFFFFFFFFFFFFFFFFFFFFFFFFFFFFFFFFFFFFF
  %[ht]
% FFFFFFFFFFFFFFFFFFFFFFFFFFFFFFFFFFFFFFFF

We find that a good fit to the points on Fig. \ref{fig:CEE_CORE_MASS_noEFFECTgraph} that we obtain from the \textsc{mesa-binary} simulations is (blue line on that figure)
\begin{equation}
 M_{\rm RGCE} (M_\odot) =
0.024 a^{0.43}_0 (R_\odot) + 0.17 .
\label{eq:dMdA}    
\end{equation}
We cast the later equation as
\begin{equation}
\log a_0 (R_\odot) = 
2.33 \log \left[ M_{\rm RGCE} (M_\odot) -0.17 \right] +3.77 ,
\label{eq:logAMehb}    
\end{equation}
from which we derive  
\begin{equation}
\frac {d \log a_0(R_\odot)}{d M_{\rm RGCE} (M_\odot)} 
= 
\frac {1.01}{\left[ M_{\rm RGCE} (M_\odot)-0.17 \right] }. 
\label{eq:dMdlogA}    
\end{equation}

\begin{figure}%[ht]
 %\centering
 \vskip -5.0 cm
 \hskip -2.00 cm
 \includegraphics[scale=0.6]{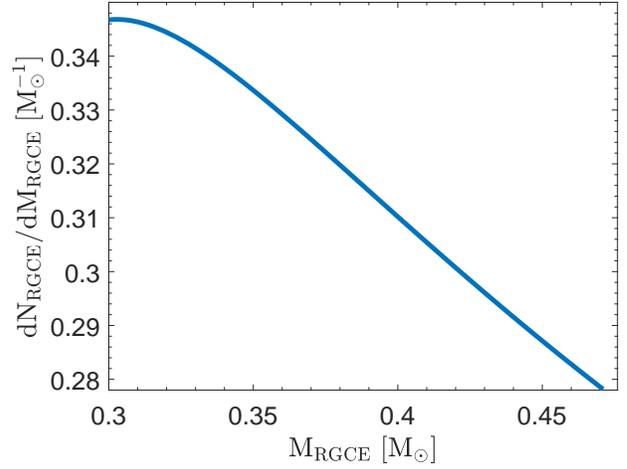}\\
 \vskip -5.00 cm
\caption{The relative number of  WD remnants of RGB CEE evolution  per mass interval, according to equation (\ref{eq:aNdMehb}), i.e., not including yet the effect of wave energy.}
 \label{fig:relative_cumulative_per_mass}
 \end{figure}

The relative number of these WDs (or to be WDs) per mass interval under our assumptions is 
 \begin{equation}
\begin{split}
& \frac {d N_{\rm RGCE}} {d M_{\rm RGCE}(M_\odot)} =   
\frac {dN}{d \log a_0(R_\odot)} 
\frac {d \log a_0(R_\odot) }{dM_{\rm RGCE}(M_\odot) } 
\\ &
\simeq \frac{  0.045 \log a_0(R_\odot)  - 0.0326 }
{\left[ M_{\rm RGCE} (M_\odot)-0.17 \right] }     
\\ & 
\simeq
\frac{0.106 \log \left[ M_{\rm RGCE} (M_\odot) -0.17 \right] +0.139 }
{M_{\rm RGCE} (M_\odot)-0.17 }
,
\label{eq:aNdMehb} 
\end{split}
\end{equation}
where we used equation (\ref{eq:dNda}) to develop $dN/d\log a_0$ and equation (\ref{eq:dMdlogA}) to develop  $d \log a_0/dM_{\rm RGCE}$. We draw relation (\ref{eq:aNdMehb}) in Fig. \ref{fig:relative_cumulative_per_mass}. 
% FFFFFFFFFFFFFFFFFFFFFFFFFFFFFFFFFFFFFFFFFFFFFFFFFFFFFFFFFFFFFFFFFFFFFFFFFFFFFFF
  %[ht]
% FFFFFFFFFFFFFFFFFFFFFFFFFFFFFFFFFFFFFFFFFFFFFFFFFFFFFFFFFFFFFFFFFFFFFFFFFFFFFFFFFFFFFFFFF

In Fig. \ref{fig:relative_cumulative_per_mass} we draw the relative number of systems from a minimum mass of $M_{\rm RGCE,min}=0.3 M_\odot$ to mass $M_{\rm RGCE}$. We chose the lower mass somewhat arbitrarily, as these small masses result form interaction with close companions when the RGB star is more compact. Therefore, its binding energy is larger and we might need to consider more massive companions to unbind the entire envelope. In any case, we are interested in the behaviour of the cumulative mass distribution near the upper limit of the mass distribution, where the EHB stars are, namely, those that ignite helium.
  
\begin{figure}%[ht]
 %\centering
 \vskip -5.0 cm
 \hskip -2.00 cm
 \includegraphics[scale=0.6]{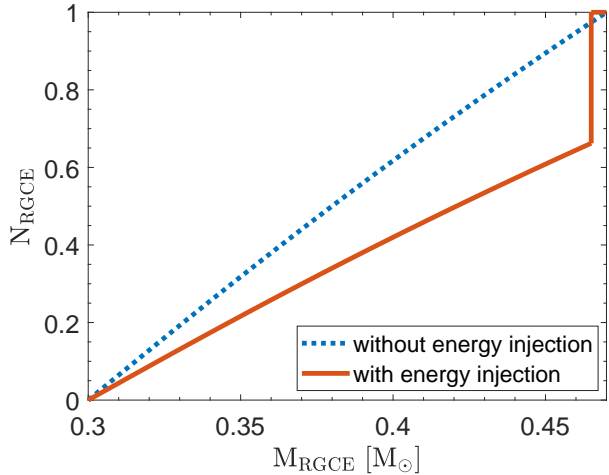}\\
 \vskip -5.00 cm
\caption{Relative cumulative mass distribution  of low-mass WDs that result from CEE on the RGB (RGB CEE).  
The blue-dotted line is according to equation (\ref{eq:Cumulative1}) when the wave energy deposition is ignored. The red-solid line is the relative cumulative mass distribution when we include the energy of the waves and assume that any secondary star that enters the briefly inflated RGB envelope removes the entire envelope. The width of the step at the upper masses is somewhat arbitrary (see text).  In this graph only stars with masses $M_{\rm RGCE} \ga 0.465 M_\odot$ are EHB stars that leave CO WD remnants. Lower mass stars leave He WD remnants.   All binaries have  $M_{\rm ZAMS,1}=1.6 M_\odot$, $M_2=0.3M_\odot$ and $e_0=0.4$.}
 \label{fig:N_EHB_vs_M_EHB}
 \end{figure}

This cumulative mass distribution is simply the integration of equation (\ref{eq:aNdMehb})
\begin{equation}
\begin{split}
& N_{\rm RGCE}(M_{\rm RGCE}) = \frac{1}{\Gamma} \int^{M_{\rm RGCE}} _{0.3M_\odot}  \frac {d N_{\rm RGCE}} {d M'_{\rm RGCE}}   d  M' _{\rm RGCE}  
\\ \simeq &  \Gamma^{-1} \{
0.0230 \ln ^2 \left[ M_{\rm RGCE} (M_\odot)-0.17 \right] 
\\ + & 
0.139 \ln \left[ M_{\rm RGCE} (M_\odot)-0.17 \right] + 0.188 \}
, 
    \label{eq:Cumulative1} 
\end{split}
\end{equation}
where $\Gamma^{-1}$ is a normalisation to give a value of $N_{\rm RGCE}(M_{\rm RGCE,max}) =1$, i.e., $\Gamma$ is the integral itself from $M_{\rm RGCE,min}=0.3 M_\odot$ to $M_{\rm RGCE,max}$, which in our simulations is $M_{\rm RGCE,max}=0.47 M_\odot$ (we checked that the same mass limit of $0.47 M_\odot$ holds also for $M_{\rm ZAMS,1}=1.2 M_\odot$). This gives $\Gamma^{-1} = 18.6$. 
We draw the cumulative mass distribution when we ignore any wave energy deposition (equation \ref{eq:Cumulative1}) with the blue-dotted line on Fig. \ref{fig:N_EHB_vs_M_EHB}.  
% FFFFFFFFFFFFFFFFFFFFFFFFFFFFFFFFFFFFFFFFFFFFFFFFFFFFFFFFFFFFFFFFFFFFFFFFFFFFFFF
  %[ht]
% FFFFFFFFFFFFFFFFFFFFFFFFFFFFFFFFFFFFFFFFFFFFFFFFFFFFFF

We also draw in Fig. \ref{fig:N_EHB_vs_M_EHB} the relative cumulative mass distribution for our simulated cases with $M_2=0.3 M_\odot$ and $e_0=0.4$, but where we do take into account the companions that enter CEE during the briefly envelope inflated phase due to deposition of wave energy to the envelope (red-solid line). In drawing this line we assume that any secondary star that enters the RGB envelope removes its entire envelope, although in reality this is not the case (section \ref{sec:PNe}). 

We calculate that cumulative mass distribution as follows. We find that, under our assumptions for $M_{\rm ZAMS,1}=1.6 M_\odot$, $M_2=0.3 M_\odot$ and $e_0=0.4$, the inflated RGB star due to the wave energy swallows secondary stars that are with initial semi-major axes in the range of  $340 R_\odot \la a_0 \la 650 R_\odot$ (Table \ref{tab:Table1}). By integrating equation (\ref{eq:dNda}) we find that the number of zero age main sequence binaries in this range is 0.47 times the number of systems in the range of $a_{\rm min}=50 R_\odot$ to $a_0=340 R_\odot$ which are the companions that enter a CEE before the core helium flash. Namely, the systems that enter a CEE before the core helium flash are now a fraction of $(1+0.47)^{-1}= 0.68$ of all cases. 
When we include all these binaries that enter a CEE under our assumption of a rapid expansion during the core helium flash we obtain the red-solid line in the Fig. \ref{fig:N_EHB_vs_M_EHB}. 
The binary systems that enter a CEE after the core helium flash leave an EHB star with a mass of $0.47 M_\odot$. However, to make the step visible and to include variations due to effects we do not include, like stellar rotation and that the secondary does not remove the entire RGB envelope (e.g., \citealt{Politanoetal2008}), we draw the step from $M_{\rm RGCE}=0.465 M_\odot$ to $M_{\rm RGCE}=0.47 M_\odot$.  These are EHB stars as they burn helium, hence they leave a CO WD remnants.  If there is some leftover envelope the distribution will continue to higher masses, $\simeq 0.54 M_\odot$ \citep{Politanoetal2008}. 

In section \ref{sec:PNe} we discuss our expectation that not all secondary stars that enter the envelope during the briefly RGB inflated phase due to wave energy deposition remove the entire envelope. Therefore, we expect that in reality, and if waves carry energy to inflate the envelope, the step at $M_{\rm RGCE} \simeq 0.47 M_\odot$ in the cumulative mass distribution will be shallower and extend to somewhat higher masses than what we present here under the assumption that all secondaries that enter a CEE remove the entire envelope. 
In any case, the red-solid line in Fig. \ref{fig:N_EHB_vs_M_EHB} should be taken as a more qualitative demonstration of the effect of waves that the core convection during core helium flash excites. The reason is the highly uncertain values of the energy that the waves deposit to the envelope and the way the energy is distributed in the envelope (see \citealt{Bearetal2021, Merlovetal2021}), and the effects of different secondary masses and retained envelope masses (e.g., population study by \citealt{Politanoetal2008}). 

% ==========================================================
\section{Implication to planetary nebulae (PNe)}
\label{sec:PNe}
% ==========================================================

Non-spherical PNe evolve from binary systems (e.g., reviews by  \citealt{DeMarco2009,Blackman2022}) and from planetary systems (e.g., \citealt{DeMarcoSoker2011}). 

Low mass solar-like stars also form PNe (e.g., \citealt{Badeneseetal2015}). 
An issue with low mass stars as progenitors of PNe is the following. Low mass stars of $0.8 \la M_{\rm ZAMS} \la 1.5 M_\odot$ have their maximum radius on the RGB not much smaller, or even larger, than their maximum radius on the asymptotic giant branch (AGB; e.g.,  \citealt{IbenTutukov1985, Nordhausetal2010}). If a star in that mass range does not engulf its companion and does not enter a CEE along its RGB phase, it is unlikely to enter a CEE during the AGB phase. The reason is that from the RGB tip to the AGB tip the star loses mass and the orbital separation increases. This by itself suggests that low mass stars of $ \la 1.5 M_\odot$ cannot form non-spherical PNe. But such stars are observed to form elliptical PNe.  
One solution to this problem is that the mass loss rate before binary interaction takes place is much lower than what canonical values are, and therefore the star reaches a much larger radius on the AGB than what canonical evolution simulations give (e.g., \citealt{SabchSoker2018a, SabchSoker2018b}). 

We here raise another solution to this problem of shaping non-spherical PNe that are descendant of low mass stars. 

Consider the evolution during the brief envelope expansion due to the energy of the waves that the core helium flash excites. The significant expansion lasts for $\simeq 2-10 \yr$ \citep{Bearetal2021}. During that time the RGB star engulfs a secondary star at an orbital separation of $\simeq {\rm several} \times 100 R_\odot$ (Fig. \ref{fig:CEE_INJECTIONgraph} here and \citealt{Merlovetal2021}). The orbital period is one to few years. As the secondary spirals-in inside the RGB envelope, the envelope shrinks because the wave energy deposition decreases.  The companion deposits orbital energy to the envelope as it spirals-in, prolonging somewhat the expanded phase of the envelope.  On the other hand, the spiral-in process removes outer envelope shells for the same reason.  
 
We raise the possibility that although the secondary spirals-in, it ends at an orbital separation of $a_{\rm f} \ga 100 R_\odot$, just near and about the RGB radius before wave energy deposition, and does not remove the entire envelope. The primary star evolves then to the horizontal branch while maintaining a massive enough envelope, $\ga 0.5 M_\odot$, to evolve along the AGB. But now the secondary stellar orbit is much smaller and the primary star manages to engulf the secondary star on the AGB. This evolutionary track leads to the formation of a non-spherical PN. 

\begin{figure}%[ht]
 %\centering
 \vskip -5.0 cm
 \hskip -2.00 cm
 \includegraphics[scale=0.6]{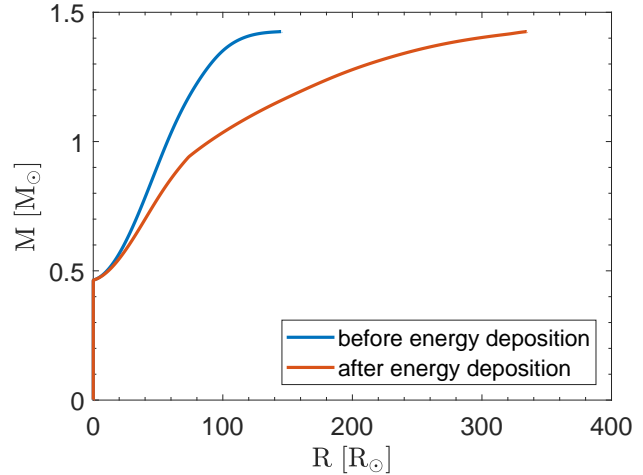}\\
 \vskip -5.00 cm
\caption{The mass as function of radius of our $M_{\rm ZAMS,1}=1.6M_\odot$ stellar model just before the core helium flash (blue line) and near maximum expansion (red line) after deposition of the wave energy that we use in this study (section \ref{subsec:Procedure}). }
 \label{fig:MvsR}
 \end{figure}

In Fig. \ref{fig:MvsR} we present the mass as function of radius near maximum expansion after wave energy deposition. We learn that there is a sufficient envelope mass at $r \simeq 100-200 R_\odot$ to cause the secondary to spiral-in, and that if the secondary ends at $a_{\rm f} \simeq 100-200 R_\odot$ it leaves sufficient mass for the primary star to evolve later along the AGB and form a PN.   
% FFFFFFFFFFFFFFFFFFFFFFFFFFFFFFFFFFFFFFFFFFFFFFFFFFFFFFFFFFFFFFFFFFFFFFFFFFFFFFF
  %[ht]
% FFFFFFFFFFFFFFFFFFFFFFFFFFFFFFFFFFFFFFFFFFFFFFFFFFFFFF
  
The assumption here is that in a large fraction of the binary systems that enter a CEE during the brief envelope expansion the primary star maintains a sufficiently massive envelope to later form a PN. We consider this to be a more realistic assumption than the assumption that in all these cases the CEE removes the entire RGB envelope--the assumption we made in deriving equation (\ref{eq:Cumulative1}) and in drawing the red-solid on fig. \ref{fig:N_EHB_vs_M_EHB}. 

Our proposed scenario to account for non-spherical PNe by low mass stars, $M_{\rm ZAMS} \la 1.5 M_\odot$, requires confirmation by conducting simulations of the CEE that results from the brief envelope expansion during the core helium flash. Such simulations require three-dimensional hydrodynamical numerical codes.  

% ==========================================================
\section{Summary}
\label{sec:Summary}
% ==========================================================

We based our study on the speculative assumption \citep{Bearetal2021, Merlovetal2021} that during core helium flash the vigorous core convection excites waves that carry energy to the envelope, and that the energy is sufficiently high to cause the envelope to inflate within few years. For the power of the waves, the duration of their activity, and the zone in the envelope where the waves deposit their energy (section \ref{subsec:Procedure}) we took intermediate values from the ranges of parameters that these earlier papers used. 

Using the open-source stellar evolutionary code \textsc{mesa-binary} (e.g., \citealt{Paxtonetal2019}) we simulated the evolution of binary systems towards a CEE along the RGB of the primary star. We found the maximum initial semi-major axes for which the binary systems enter CEE along the RGB before core helium flash (third column of Table \ref{tab:Table1}) and during the core helium flash when waves deposit energy to the envelope according to our assumptions (fourth column of Table \ref{tab:Table1}). We presented  the evolution of orbital parameters for one example from each of these two types of CEE onset in Fig. \ref{fig:CEE_NO_INJECTIONgraph} and Fig. \ref{fig:CEE_INJECTIONgraph}, respectively. 

We discussed two possible implications of the process by which flash-excited waves carry energy from the convective core to the RGB envelope and cause a brief (several years) envelope inflation.  

In section \ref{sec:Cumulative} we argued that the process increases the number of EHB (sdBO) stars. We demonstrated the effect by considering binary systems with initial parameters of $M_{\rm ZAMS,1}=1.6M_\odot$, $M_2=0.3M_\odot$ and $e_0=0.4$. Without the effect of the waves the expected cumulative mass distribution of WDs that result from RGB CEE is approximately as the blue-dotted line in Fig. \ref{fig:N_EHB_vs_M_EHB} presents. If all secondary stars that enter a CEE during the core helium flash as a result of the rapid envelope inflation remove the entire envelope, on the other hand, the cumulative mass distribution is as given by the red-solid line in Fig. \ref{fig:N_EHB_vs_M_EHB}. The width of the step is arbitrary, as it should be very small by our assumptions (see section \ref{sec:Cumulative}). 
 
However, we expect that only in a  small  fraction of the cases where the binary system enters a CEE during the core helium flash the secondary manages to   actually  remove the entire envelope. Therefore, we expect the cumulative mass distribution to be closer to blue-dotted line than to the red line of Fig. \ref{fig:N_EHB_vs_M_EHB}. As well, we expect the step to be wider and to extend to somewhat higher masses. 
In any case, we predict a sharp,  although small,  peak in the cumulative mass distribution at its upper mass edge where EHB stars are, $M_{\rm RGCE} \simeq 0.47 M_\odot$. We note that \cite{Politanoetal2008} also predict a peak around this mass range using a population synthesis study and without the effect of waves. The shape of the peak in our study is different than in their study, but we did not include population synthesis. Future studies should incorporate the effects of waves into population synthesis calculations. 

 Overall, in most cases we do not expect the companion to remove the entire envelope when the core helium flash causes the CEE on the tip of the RGB. For that, we predict that the proposed envelope expansion during the core helium flash increases the expected number of EHB stars by no more than several tens of percents. We cannot at this stage be more accurate. On the other hand, the leftover envelope after the CEE due to core helium flash implies that in most cases the star later forms a planetary nebula, but now the companion is closer to the star than what evolution without envelope expansion at core helium flash predicts.  

In section \ref{sec:PNe} we raised the possibility that the effect of the flash-excited waves might ease the formation of non-spherical PNe by low mass stars. The radii of low mass stars, $M_{\rm ZAMS} \la 1.5 M_\odot$, as they reach the tip of their AGB are not much larger, or even smaller, than their radii on the RGB tip. Therefore, in most cases of canonical stellar evolution if the primary star does not engulf the secondary star on the RGB it will neither engulf it on the AGB. This crucially depend on the mass loss rate though. We suggested that in some cases the rapid expansion of the RGB star due to the energy deposition by the flash-excited waves might trigger a CEE but for a short time of less than few years, during which the secondary star spirals-in but to a large radius of $a_{\rm f} \simeq 100-200 R_\odot$, and does not remove the entire envelope. The leftover envelope allows the primary star to evolve later along the AGB, engulf the secondary star, and form a non-spherical PN. Future studies should examine this short-duration CEE following the rapid RGB expansion with three-dimensional hydrodynamical simulations.

%%%%%%%%%%%%%%%%%%%%%%%%%%%%%%%%%%%%%%%%%%%%%%%%
%\acknowledgments
\section*{Acknowledgements}
%%%%%%%%%%%%%%%%%%%%%%%%%%%%%%%%%%%%%%%%%%%%%%%%
We thank Matthias Kruckow for useful comments,  and an anonymous referee for useful comments and corrections.  This research was supported by a grant from the Israel Science Foundation (769/20).  

%%%%%%%%%%%%%%%%%%%%%%%%%%%
\section*{Data Availability}

The data underlying this article will be shared on reasonable request to the corresponding author. 
%%%%%%%%%%%%%%%%%%%%%%%%%%%

%\software{MESA \citep{Paxtonetal2011, Paxtonetal2013, Paxtonetal2015,Paxtonetal2018, Paxtonetal2019}}

\pagebreak

\end{document}